\documentclass[aps,prl,twocolumn]{revtex4-2}

\usepackage{graphicx}
\usepackage{mathrsfs}

\def\E{\tilde{E}}

\def\d{\rm d}

\def\B{\mathscr{B}}

\begin{document}


\title{Spatial cage solitons --- taming light bullets}


\author{Chao Mei}
\affiliation{Max Born Institute for Nonlinear Optics and Short Pulse Spectroscopy, 12489 Berlin, Germany\\
 and School of Computer and Communication Engineering, University of Science and Technology Beijing (USTB), 100083 Beijing, China}

\author{Ihar Babushkin}
\affiliation{Institute of Quantum Optics, Leibniz University Hannover, 30167 Hannover, Germany}

\author{Tamas Nagy}
\affiliation{Max Born Institute for Nonlinear Optics and Short Pulse Spectroscopy, 12489 Berlin, Germany}

\author{Günter Steinmeyer}
\email{steinmey@mbi-berlin.de\\ }
\affiliation{Max Born Institute for Nonlinear Optics and Short Pulse Spectroscopy,  12489 Berlin, Germany\\
 and Institut f\"ur Physik, Humboldt Universit\"at zu Berlin,  12489 Berlin, Germany}



\date{\today}

\begin{abstract}
Multimode nonlinear optics offers to overcome a long-standing limitation of fiber optics, tightly phase locking several spatial modes and enabling the coherent transport of a wavepacket through a multimode fiber. A similar problem is encountered in the temporal compression of multi-mJ pulses to few-cycle duration in hollow gas-filled fibers. Scaling the fiber length to up to six meters, hollow fibers have recently reached 1\,TW of peak power. Despite the remarkable utility of the hollow fiber compressor and its widespread application, however, no analytical model exists to enable insight into the scaling behavior of maximum compressibility and peak power. Here we extend a recently introduced formalism for describing mode-locking to the spatially analogue scenario of locking spatial fiber modes together. Our formalism unveils the coexistence of two soliton branches for anomalous modal dispersion and indicates the formation of stable spatio-temporal light bullets that would be unstable in free space, similar to the temporal cage solitons in mode-locking theory. Our model enables deeper understanding of the physical processes behind the formation of such light bullets and predict the existence of multimode solitons in a much wider range of fiber types than previously considered possible.
\end{abstract}


\maketitle

\markboth{Chao Mei \textit{et al.} \hfill {\bf Spatial Cage Solitons --- Taming Light Bullets} \hfill}{Chao Mei \textit{et al.} \hfill {\bf Spatial Cage Solitons --- Taming Light Bullets} \hfill}

Spatial solitons have fascinated researchers since the early days of nonlinear optics \cite{Townes,Zakharov}. In combination with self-phase modulation, the self-focusing effect offers the possibility for three-dimensional contraction of an optical wavepacket and concomitant intensity increase. While there exist numerous reports on such light bullets \cite{Silberberg,Moll,Dubietis}, this intriguing nonlinear mechanism found little application, probably because of the limiting action of a spatial modulation instability \cite{Bespalov,Bliss}. Starting from small imperfection in the beam profile, this process induces a rapid small-scale breakup of the beam profile into filaments when the critical power $P_{\rm crit}$ is exceeded \cite{Marburger,Boyd}, thus limiting the obtainable nonlinear interaction length. Given this severe constraint, high power pulse compression and nonlinear conversion techniques have resorted to hollow capillaries for extended nonlinear interaction length \cite{MarcatiliSchmeltzer,Nisoli,Nisoli2,Durfee}. While several other competing techniques \cite{selfcompress,multipass} have been discussed for the compression of pulses with gigawatt peak powers, the hollow fiber is currently the most established compression technique and found widespread application in attosecond pulse generation and other high-field experiments, see, e.g., \cite{Sansone,Ma}. Utilizing the advanced stretched-fiber technique \cite{stretchedfiber,Nagy} recently enabled record-breaking continuous powers above 300\,W \cite{Nagy2} and peak powers exceeding 1\,TW for the first time \cite{Nagy3}. Despite the widespread utility of this technique, however, there exist only relatively few analytical approaches \cite{Tempea,Fibich} for modeling the nonlinear broadening processes inside the hollow fiber. Numerical simulations often resorted to a simplified one-dimensional approximation as full modal expansions \cite{Nurhuda,Kolesik,Bruno} are numerically cumbersome. In the following, we present a completely analytical approach for determining spatial soliton solutions in nonlinear multimode fiber geometries. Similar spatio-temporal solitons have previously been observed in numerical simulations \cite{Crego,Continuum}. Moreover, our approach is mathematically similar to the cage soliton solutions of the Haus master equation of mode-locking \cite{Haus,CageSoliton}, and it is also applicable for the thriving field of multimode fiber nonlinear optics \cite{Wise,Wise2,Christodoulides}. Assuming adiabatic pulse shaping, the results of this analysis enable the derivation of universal scaling laws for the design of nonlinear multimode waveguides.

\begin{figure*}[tb]
\centering
\includegraphics[width=0.6 \linewidth]{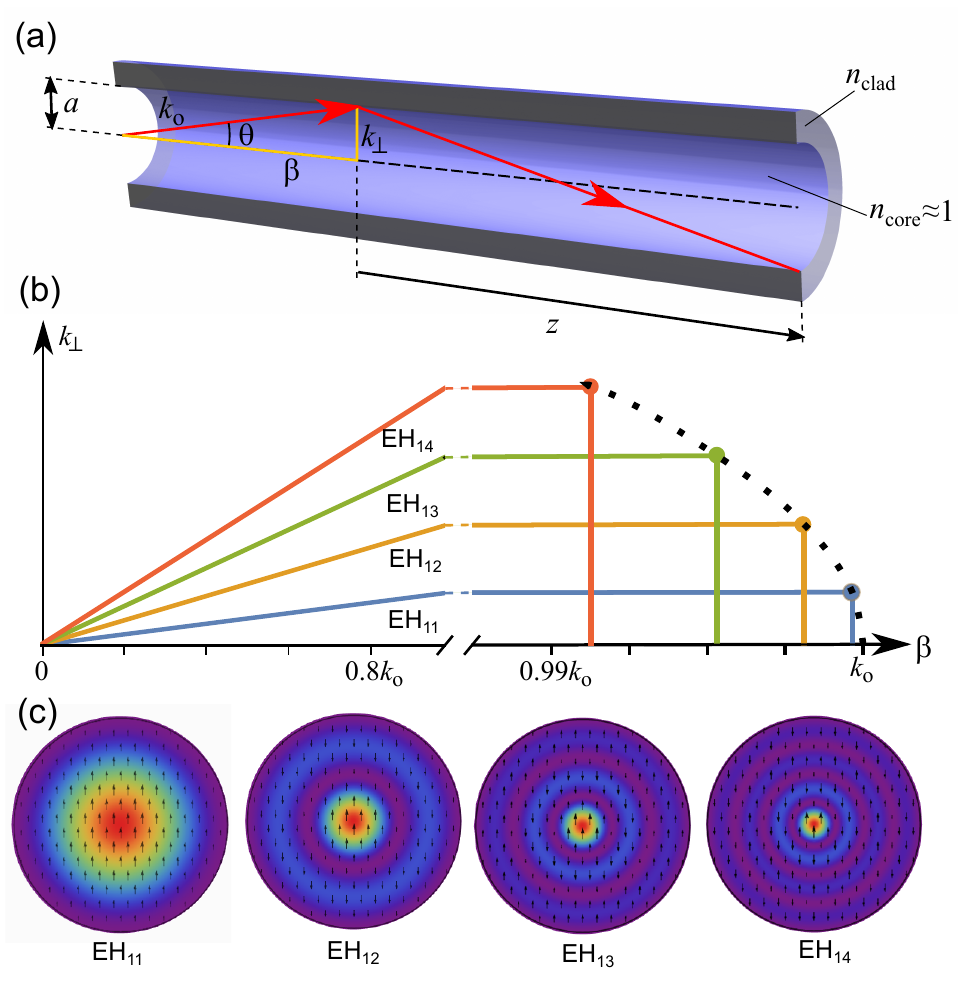}
\caption{(a) Ray-optical representation of hollow-fiber transmission. The wavevector $k_0$ can be decomposed into a transverse component $k_\perp$ and a longitudinal component $\beta$, which are connected by Pythagoras' Theorem \cite{Crenn}. (b) The propagation constants $\beta_n$ of the individual EH$_{1n}$ modes follow an approximate $n^2$ dependence whereas the $k_n^{\perp}$ underly a linear relationship with $n$. (c) Mode fields of the first four EH$_{1n}$ modes considered in this study. Intensities are depicted by colors and electric fields are represented by arrows.}
\label{fig:intro}
\end{figure*}

The linear optical properties of a cylindrical hollow dielectric waveguide with radius $a$ have been first modeled by Marcatili and Schmeltzer \cite{MarcatiliSchmeltzer}. Assuming linear polarization, one finds hybrid solutions of the wave equation, which have originally been designated as ${\rm EH}_{mn}$ modes. In solid core multimode (SCM) fibers, widely similar solutions are referred to as ${\rm LP}_{(m-1)n}$ modes. Strictly speaking, both designations are not identical because of the Goos-Hänchen effect \cite{GoosHaenchen}, which causes the ${\rm LP}_{(m-1)n}$ modes to extend into the cladding. In contrast, the ${\rm EH}_{mn}$ exhibit a node of the electric field at the dielectric interface. Assuming $m=1$, i.e., azimuthal homogeneity, the radial field profile of the ${\rm EH}_{0n}$ mode is given by
\begin{equation}
E_{n}(r) \propto J_0 \left( \frac{u_{n} r}{a} \right),
\end{equation}
with the zero-order Bessel function $J_0$ and its $n$th zero $u_{n}$ ($n>0$). The complex-valued propagation constant of these modes is given by
\begin{equation}
\kappa_{n} = \beta_n + i \alpha_n = k_0 - \frac{u_{n}^2}{2 k_0 a^2} \left(1 - i \frac{\epsilon+1}{k_0 a \sqrt{\epsilon-1}} \right),
\end{equation}
where $k_0 = 2 \pi / \lambda$ is the wavenumber and $\sqrt{\epsilon}=n_{\rm clad}$ the refractive index of the cladding material. Exploiting the trigonometric relationship between propagation constant and wave number displayed in Fig.~\ref{fig:intro}, one finds an approximate parabolic dependence \cite{Crenn,Crenn2}
\begin{eqnarray}
 \beta_n =  k_0 \cos \left( \frac{u_n}{k_0 a} \right) & \approx & \beta_1 - \frac{ \pi^2 (n-1)^2}{2 k_0 a^2} \nonumber \\
  & = & \beta_1 + \B (n-1)^2. \label{eq:para}
\end{eqnarray}
This approximation may appear crude for small $n$ and mostly serves to point out the close analogy to the role of dispersion in the propagation of femtosecond pulses. Therefore, all subsequent computations have been repeated with and without the approximation, yet with little resulting difference in the outcome. Moreover, as $\beta_n<k_0$ in a hollow waveguide, the modes $E_n$ propagate at superluminal phase velocity inside, which can be explained by $-\pi$ phase jumps due to Fresnel reflection at the interface in the geometrical picture of Refs.~\cite{Crenn,Crenn2}. A similar parabolic approximation can be made for SCM fibers \cite{LoveSnyder}, yet with positive curvature $\B$. Exploiting the relation
\begin{equation}
\beta^2+k_\perp^2 = k_0^2 \sin^2 \theta + k_0^2 \cos^2 \theta = k_0^2
\end{equation}
depicted in Fig.~\ref{fig:intro}(a), one also finds an approximate linear relationship for the transverse wave number
\begin{equation}
k_n^{(\perp)} = k_0 \sin \left( \frac{u_n}{k_0 a} \right) \approx \left(n-\frac{1}{4}\right)\frac{\pi}{a}.  \label{eq:lin}
\end{equation}
In the following, we describe the evolution of the spatial beam profile $E(z,r)$ upon propagation along the coordinate $z$ with the transverse wave equation
\begin{equation}
\partial_z E = \frac{i \beta}{r} \partial_r r \partial_r E + i \, \Gamma |E|^2 E,   \label{eq:PDE}
\end{equation}
due to competing diffractive effects and the action of self-focusing with $\Gamma=k_0 n_2$ and the nonlinear refractive index $n_2$. Equation (\ref{eq:PDE}) is widely similar to the Nonlinear Schrödinger Equation (NLSE), yet with the temporal coordinate $t$ replaced by the radial coordinate $r$. While it is customary to numerically solve the NLSE by a split-step Fourier method \cite{Agrawal}, the equation can be reformulated in the frequency domain \cite{CageSoliton}, which leads to a system of nonlinearly coupled ordinary differential equations. Rewriting Eq.~(\ref{eq:PDE}) in the $k^{(\perp)}$ domain leads to a similar system of coupled equations \cite{Crego}
\begin{equation}
\partial_z \tilde{E}_n = i \beta_n \tilde{E}_n + i \, \Gamma \mkern-18mu \sum_{j+k-\ell=n} \mkern-18mu \tilde{E}_j \tilde{E}_k \tilde{E}^*_\ell .  \label{eq:ODE1}
\end{equation}
Here the summation over the four-wave mixing terms is restricted to those products that satisfy conservation of the transverse wavenumber $k_\perp$, cf.~Fig.\ref{fig:intro}(b). The partially degenerate case $j=k$ is also known as self-diffraction \cite{selfdiffract}. As was noted by DeLong \textit{et al.}~\cite{Trebino}, the self-diffraction process is not exactly phase matched. This slight mismatch is accounted for by the intermodal dispersion term $\beta_n \tilde{E}_n$, which takes the part of group-delay dispersion with its quadratic dependence in the frequency representation of the Haus master equation \cite{CageSoliton}. As the parabolic dependence on mode number $n$ is convex for hollow fibers one can associate this case with anomalous dispersion whereas SCM fibers display normal modal dispersion. Moreover, the slight phase mismatch can be understood with the neglection of the linear term in Eq.~(\ref{eq:para}) and is otherwise similar to higher-order dispersion contributions in traditional mode-locking theory \cite{Haus}, i.e., these contributions give rise to deviations from equidistance of the cold cavity modes. Solving the mode-locking version of Eqs.~(\ref{eq:ODE2}) and (\ref{eq:algebra}) requires restrictive assumptions on the number of coupled longitudinal modes \cite{CageSoliton}. In the transverse case, it is often considered sufficient to include only a few spatial modes (e.g., $N=3$ \cite{Bruno}) for treating nonlinear propagation through a hollow waveguide. We therefore write out Eq.~(\ref{eq:ODE1}) for 4 modes, using excessive content in $\E_4$ as an indicator for the breakdown of our simplifying assumptions
\begin{widetext}
\begin{eqnarray}
\partial_z \E_1 & = & i \beta_1 \E_1 + i \Gamma \left[\eta_{1234} \E_2 \E_3 \E_4^*+ \eta_{123} \E_2^2 \E_3^* + \left(|\E_1|^2+ \eta_{12} |\E_2|^2+ \eta_{13} |\E_3|^2+ \eta_{14} |\E_4|^2 \right) \E_1 \right] \nonumber \\
\partial_z \E_2 & = & i \beta_2 \E_2 + i \Gamma \left[\eta_{1234} \E_1 \E_3^* \E_4 + \eta_{123} \E_1 \E_2^* \E_3 +  \eta_{234} \E_3^2 \E_4^*  \right. \nonumber \\
& & + \left. \left( \eta_{12} |\E_1|^2 + |\E_2|^2 +  \eta_{23} |\E_3|^2 + \eta_{24}|\E_4|^2 \right) \E_2 \right] \nonumber \\
\partial_z \E_3 & = & i \beta_3 \E_3 + i \Gamma \left[ \eta_{1234} \E_1 \E_2^* \E_4  + \eta_{123} \E_1^* \E_2^2 + \eta_{234} \E_2 \E_3^* \E_4  \right. \nonumber \\
 &  & + \left. \left( \eta_{13} |\E_1|^2 + \eta_{23} |\E_2|^2+ |\E_3|^2 + \eta_{34} |\E_4|^2 \right) \E_3 \right] \nonumber \\
\partial_z \E_4 &=& i \beta_4 \E_4 + i \Gamma  \left[ \eta_{1234}  \E_1^* \E_2 \E_3 + \eta_{234} \E_2^* \E_3^2 \left( \eta_{14} |\E_1|^2 + \eta_{24} |\E_2|^2 + \eta_{34} |\E_3|^2 + |\E_4|^2 \right) \E_4 \right]. \nonumber \\
\label{eq:ODE2}
\end{eqnarray}
\end{widetext}
Here the $\beta_n$ terms describe intermodal dispersion, i.e., the phase velocity differences between the individual modes. Self-diffraction is accounted for by partially degenerate four-wave mixing terms $\propto \eta_{k \ell} \E^2_k E^*_\ell$. Self-focusing of the individual modes is described by the fully degenerate terms $\propto |E_k|^2 E_k$. In addition, cross-phase modulation (XPM) terms appear, cf.~Table 1. In Eq.~(\ref{eq:ODE2}), we introduced modal overlap factors $\eta_{nj}$, $\eta_{njk}$ and $\eta_{njk\ell}$ for non-collinear four-wave mixing processes as they have been previously discussed, e.g., by Chapman \textit{et al.} \cite{Chapman}. For the fully non-degenerate process we define
\begin{equation}
\eta_{njk\ell} = \frac{\int\limits_0^a \E_n \E_j \E_k \E_\ell r {\rm d}r}{\prod\limits_{m=\{n,j,k,\ell\}} \sqrt[4]{\int_0^a \E_m^4 r {\rm d}r}}.
\end{equation}
For degenerate mixing processes, we use a shorthand notation, e.g., $\eta_{1221}=\eta_{2112}=\eta_{12}/2$ and include the degeneracy factor in the respective overlap factors. Values for the various $\eta$ are listed in Table \ref{tab:overlap}. Provided that the nonlinear length $L_{\rm NL}$ is much shorter than the dispersive length $L_{\rm D}$ \cite{Agrawal}, Eq.~(\ref{eq:ODE2}) can be used as a highly efficient tool for simulating the propagation via solving a set of coupled ordinary differential equations. As will be further discussed below, such adiabaticity can be assumed in typical hollow-fiber compression scenarios. Moreover, propagation losses can be accounted for by a complex-valued redefinition of the $\beta_n$.

Assuming propagation of the $\E_j$ at identical phase velocity, i.e., as solitonic wavepacket, we find a wavenumber offset $\psi$ relative to the fundamental mode $\beta_1$. We further renormalize real-valued electric field amplitudes $a_n=|\E_n / \E_1|$ to yield $a_1=1$ and redefine an effective nonlinearity $\gamma=\Gamma \B^{-1} |\E_1|^{-2}$. Using these simplifications, we can extract an algebraic discriminant for the resulting spatial soliton
\begin{widetext}
\begin{eqnarray}
\psi & = & \gamma \left( 1 + \eta_{1234} a_2 a_3 a_4 + \eta_{123} a_2^2 a_3 + \eta_{12} a_2^2+ \eta_{13} a_3^2+ \eta_{14} a_4^2 \right) \nonumber \\
 & = & \gamma \left[ \eta_{1234} a_3 a_4 + \eta_{123} a_2 a_3 +  \eta_{234} a_3^2 a_4  + \left( \eta_{12} + a_2^2 +  \eta_{23} a_3^2 + \eta_{24} a_4^2 \right) a_2 \right] + a_2  \nonumber \\
 & = & \gamma \left[ \eta_{1234} a_2 a_4  + \eta_{123} a_2^2 + \eta_{234} a_2 a_3 a_4 + \left( \eta_{13} + \eta_{23} a_2^2+ a_3^2 + \eta_{34} a_4^2 \right) a_3 \right] + 4  a_3 \nonumber \\
 &=& \gamma \left[ \eta_{1234}  a_2 a_3 + \eta_{234} a_2 a_3^2 +  \left( \eta_{14} + \eta_{24} a_2^2 + \eta_{34} a_3^2 + a_4^2 \right) a_4 \right] + 9 a_4 .
\label{eq:algebra}
\end{eqnarray}
\end{widetext}

\begin{table}[tb]
\centering
\caption{Nonlinear mode coupling factors.}
\begin{tabular}{|l|ccc|ccc|ccc|}
\hline
XPM & $\eta_{12}$ & = & 1.709 &  $\eta_{13}$ & = & 1.529 & $\eta_{14}$ & = & 1.408 \\
XPM & $\eta_{23}$ & = & 1.847 &  $\eta_{24}$ & = & 1.725 & $\eta_{34}$ & = & 1.897 \\
FWM & $\eta_{123}$ & = & 0.788 &  $\eta_{234}$ & = & 0.874 & $\eta_{1234}$ & = & 1.462 \\
\hline
\end{tabular}
\label{tab:overlap}
\end{table}

One can now retrieve all real-valued roots of Eq.~(\ref{eq:algebra}) and compute beam diameter $w_{\rm eff}$, loss $\alpha_{\rm eff}$, and, most importantly, the resulting ``soliton phase'' $\psi$ \cite{solitonphase}, which is more correctly defined as the total propagation constant involving both linear and nonlinear effects relative to the phase velocity frame. $\psi$ vanishes for $\gamma \rightarrow 0$, as we already accounted for linear propagation effects by subtracting the propagation constant $\beta_1$ in  Eq.~(\ref{eq:algebra}). This effective subtraction of the linear propagation phase now gives immediate access to the nonlinear phase via $\partial\varphi_{\rm nl}/\partial z=\psi$. Using the relation $\varphi_{\rm nl}= k_0 z n_2 P A_{\rm eff}^{-1}$ with the nonlinear refractive index $n_2$ and the effective area \cite{Agrawal} of the fundamental EH$_{\rm 11}$ mode
\begin{equation}
A_{\rm eff} = \frac{1}{2 \pi} \frac{\left( \int | E(r) |^2 r \d r \right)^2 }{\int | E(r) |^4 r \d r} \approx 1.50 a^2,
\end{equation}
one can then relate the soliton phase to the power $P$ in the nonlinear waveguide. Inserting the definition of the critical power for self-focusing $P_{\rm cr}=0.147 \lambda^2/n_2$ \cite{Boyd,Marburger}, we then finally yield
\begin{equation}
P \approx 10.2 P_{\rm cr} \frac{\psi a^2}{k_0 \lambda^2}.
\end{equation}

\begin{figure*}[tb]
\centering
\includegraphics[width=0.57 \linewidth]{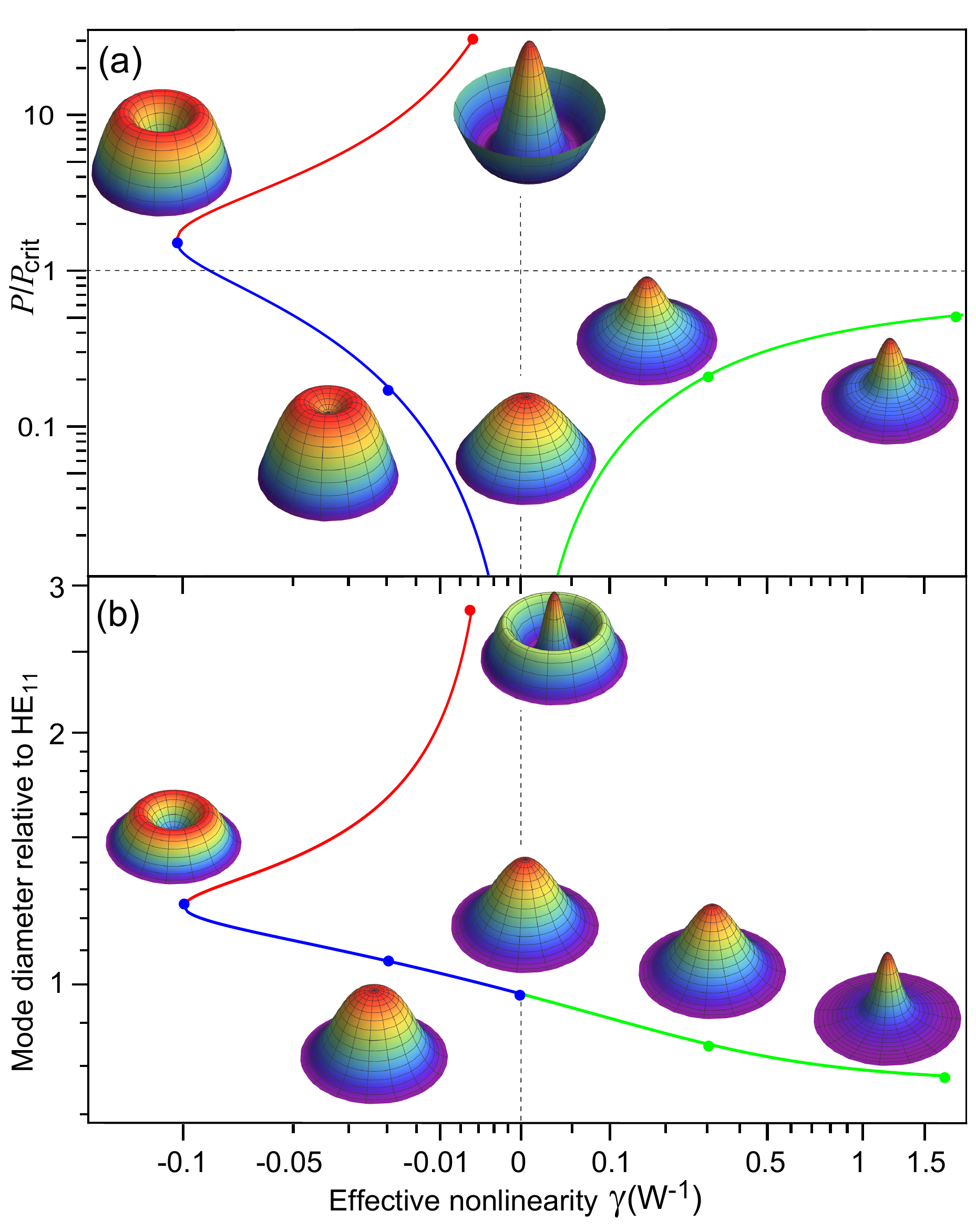}
\caption{Spatial soliton solution branches of Eq.~(\ref{eq:ODE2}). For normal modal dispersion ($n_{\rm core}>n_{\rm clad}$), a single solution branch exists (green). In hollow fibers, two branches coexist (blue and red). The red branch is considered unstable (see discussion in text). (a) Radially integrated intensity $\int E(r)^2 r{\d}r=P$ of the spatial cage soliton solutions  vs. effective nonlinearity. Powers have been normalized to the critical power of self-focusing $P_{\rm cr}$ in free space \cite{Boyd}. Insets show $E(r)$ for parameters indicated by symbols. (b) Root mean square mode diameter of spatial solitons normalized to the HE$_{11}$ mode. Insets show spatial intensity profiles $|E(r)|^2$.}
\label{fig:2}
\end{figure*}
\begin{figure}[tb]
\centering
\includegraphics[width=\linewidth]{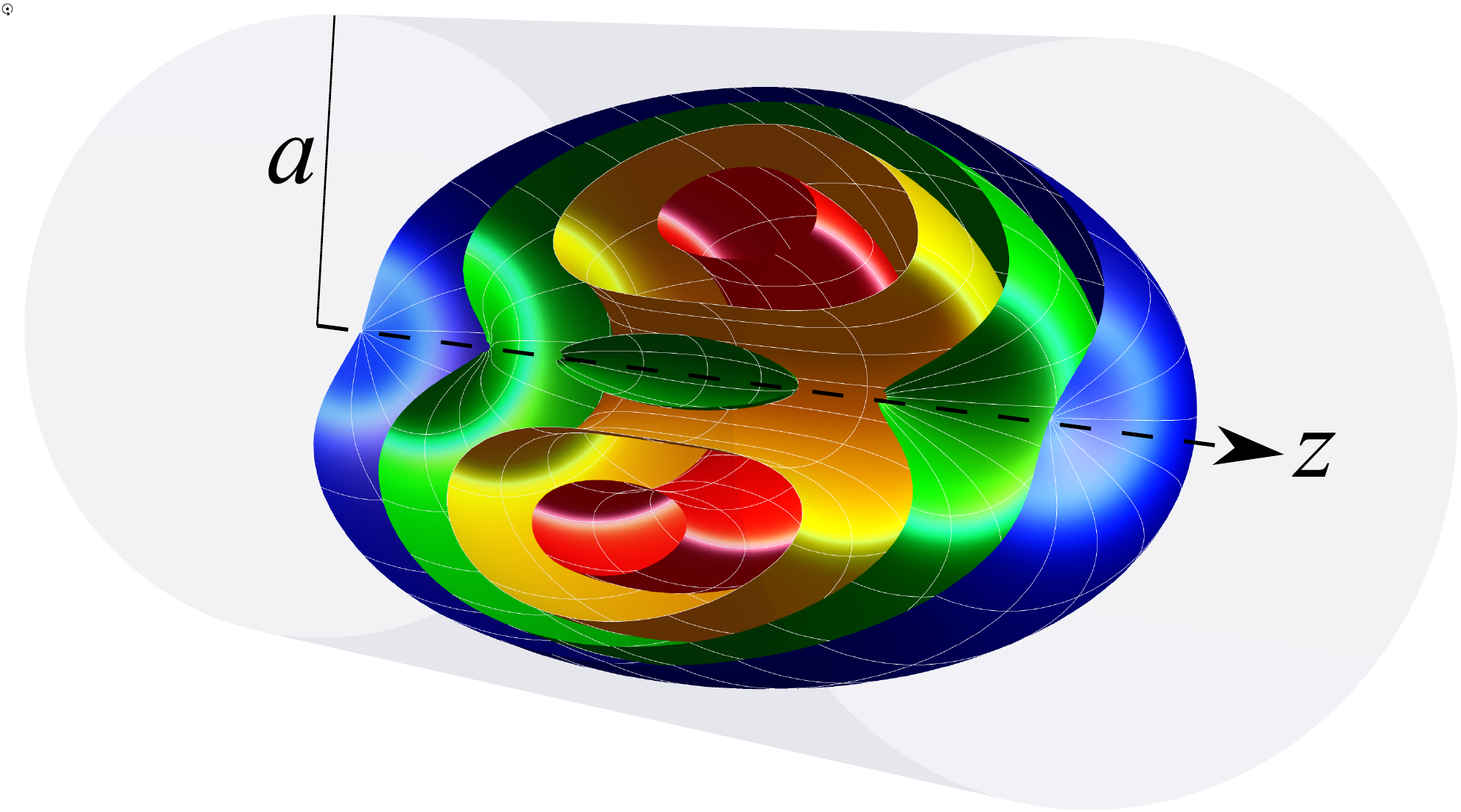}
\caption{Three-dimensional visualization of spatio-temporal soliton structure at the stability limit ($\approx 1.4 P_{\rm crit}$) in the anomalous modal dispersion regime. Equi-intensity surfaces are shown with colors red ($80\%$ peak int.) to blue ($10\%$ peak int.). In the center, a donut structure dominates which evolves into an ellipsoidal shape in the temporal wings. The glass-gas interface of the hollow fiber is depicted in light gray for comparison.}
\label{fig:4}
\end{figure}

The resulting peak power vs.\ effective nonlinearity is depicted in Fig.~\ref{fig:2}(a). In the anomalous regime (hollow fibers), the higher-order modes ${\rm EH}_{1n}, n>1$ of the spatial soliton solutions carry the opposite sign of the fundamental ${\rm EH}_{11}$ mode. This can be understood by a partial cancellation of nonlinearity and modal dispersion, similar to the formation of Schrödinger solitons \cite{Zakharov}. In turn, a depression of the peak intensity and an increased effective area $A_{\rm eff}$ of the beam profile result. [Fig.~\ref{fig:2}(b)]. Consequently, the hollow waveguide can host beams with a peak power of $\approx 1.4 P_{\rm cr}$, yet at a characteristic donut profile, cf.~Fig.~\ref{fig:4}. In the normal dispersion regime (SCM fibers), peak intensities are enhanced as nonlinearity and modal dispersion add up, leading to cusp-like beam profiles for $\kappa>0$. Comparing both waveguide dispersion regimes, it is striking that the asymmetry in the $P(\gamma)$ relation is only caused by beam profile variations. For the SCM case, this effect leads to a maximum peak power hosting of 0.5~$P_{\rm cr}$. Apart from these two fundamental soliton branches, our investigations identifies a second solution branch in the anomalous dispersion regime (red curves in Fig.~\ref{fig:2}). In this branch, the limit-value donut solution converges toward a ``Mexican hat'' solution upon subsequent reduction of power. As the latter solutions involve an increasing amount of higher-order modes, such beam profiles would experience rapidly increasing losses, i.e., this solution branch is considered unstable. In contrast, the fundamental soliton branches (blue and green in Fig.~\ref{fig:2}) self-stabilize upon propagation as their higher-order contents reduces with decreasing peak power.

To this end, it appears illustrative to compute the relevant interaction lengths \cite{Agrawal} inside a hollow fiber. Given the rather low group-velocity dispersion of noble gases like argon, which may be additionally  cancelled out by the waveguide (group-velocity) dispersion of the hollow fiber, dispersion lengths $L_D \approx \tau^2 / |\beta_2|$ for pulse durations $\tau_0>20\,$fs generally exceed the fiber length by a large factor for experimental conditions in \cite{Nagy,Nagy2,Nagy3}. The absorption length is commonly also chosen longer than the actual fiber length. This is contrasted by the nonlinear length $L_{\rm NL}=\Gamma^{-1}P^{-1}$, which amounts to only a few millimeters for powers $P > 0.1 P_{\rm cr}$ at identical experimental conditions. One can therefore conclude that spatial soliton effects strongly dominate the nonlinear dynamics inside the hollow fiber, causing an adiabatic reshaping of the solitons as a reaction to the comparatively slow waveguide losses. Non-solitonic contents is stripped off into linearly propagating higher-order modes, which travel at reduced group velocities. For example, for $a=200\,\mu$m and $\lambda=1\,\mu$m, the linear group delay of the ${\rm EH}_{12}$ relative to the fundamental mode amounts to about 100\,fs per meter propagation length, i.e., the non-solitonic contents will lead to the formation of a temporal continuum background after recompression. Ignoring reshaping effects due to group-velocity dispersion, one can now compute the structure of the emerging spatio-temporal light bullets, see Fig.~\ref{fig:4}. Here we have chosen the highest possible peak power in Fig.~\ref{fig:2}, i.e., $P=1.4\,P_{\rm cr}$, which lead to the formation of a donut spatial structure at pulse center. At lower intensities this structure goes over into the more common ellipsoidal shape of conventional light bullets.

\begin{figure}[htbp]
\centering
\includegraphics[width=\linewidth]{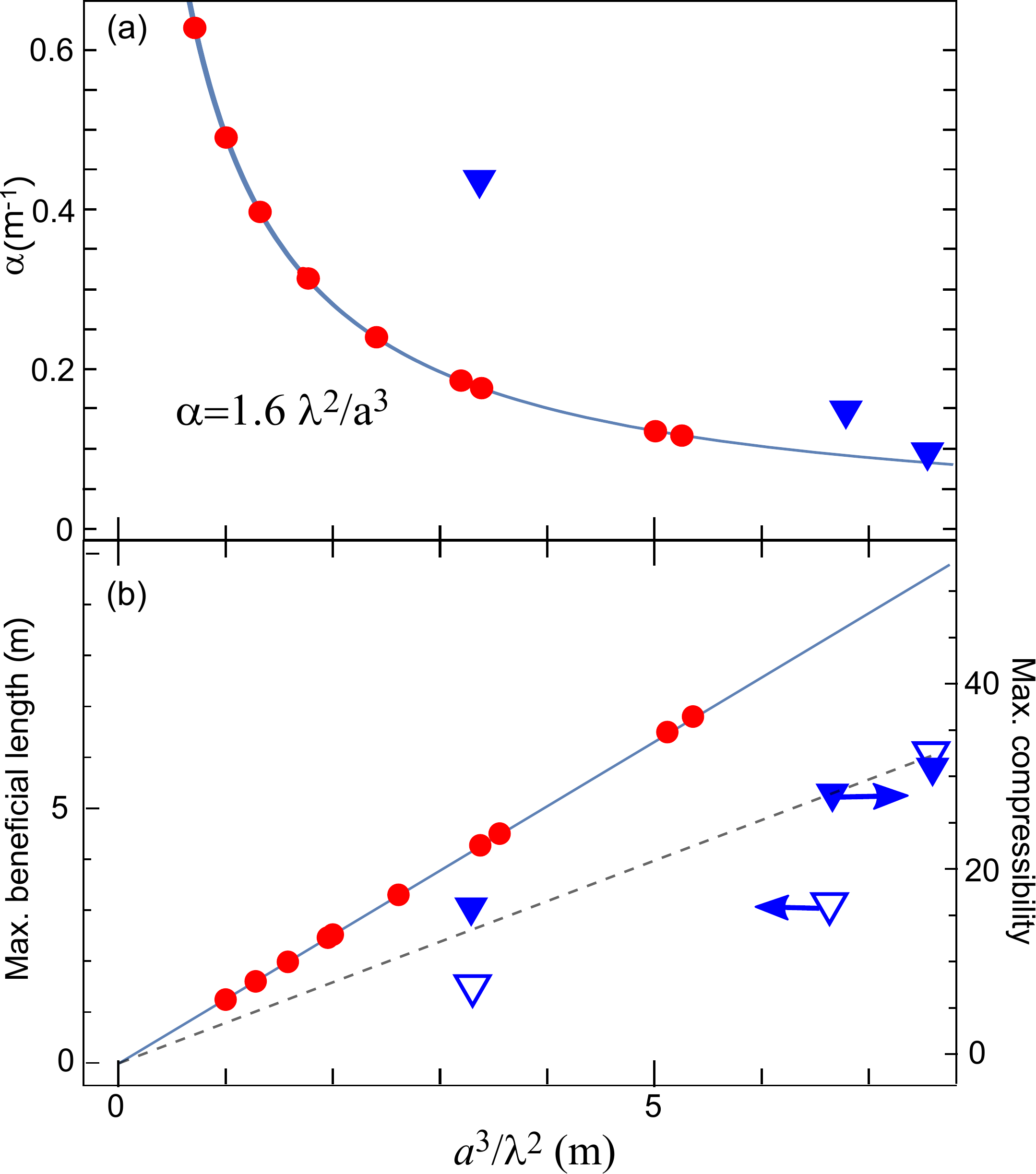}
\caption{Comparison of model results  with measured data. (a) Total losses (linear and nonlinear) vs. ratio of $a^3$ and $\lambda^2$ (curve and red dots). Early measurements with relatively short hollow fibers \cite{Nagy} indicated significantly higher losses whereas more recent measurements \cite{Nagy2,Nagy3} showed excellent agreement (blue triangles). (b) Maximum beneficial length (solid curve and solid symbols) and maximum compressibility (dashed line and hollow symbols), cf.~Eq.~(13). This analysis confirms that superior compression can be reached with longer hollow fibers and larger core diameters.}
\label{fig:3}
\end{figure}
Utilizing the spatial soliton solutions of Eq.~(\ref{eq:ODE2}), one can now derive a few design rules for hollow fiber compressors. As nonlinearly broadened spectra typically exhibit near-perfect spectral symmetry, one can employ the simple relation \cite{Agrawal}
\begin{equation}
\chi(z)=\frac{\Delta\nu_{\rm rms}}{\Delta\nu_0}=\sqrt{1+\frac{4}{3 \sqrt{3}} \varphi_{\rm nl}^2(z)}
\end{equation}
to estimate the compressibility of the input pulse. In view of applications, increase of peak power is typically considered more important than ultimate shortness of the pulse. Let us therefore define the figure of merit
\begin{equation}
M(a,\lambda)=\left.{\rm max}\left[ \chi(z) \exp(-2 \alpha_{\rm tot}z)\right]\right|_z
\end{equation}
as the criterion for the maximum beneficial propagation length $z_{\rm max}$ inside the hollow fiber. Neglecting any possible loss in the subsequent compression process (e.g., by chirped mirrors), peak powers will not further increase upon additional propagation. Pertinent computations are shown in Fig.~\ref{fig:3}, indicating the dependence of the total (linear and nonlinear) loss $\alpha_{\rm tot}$ as a function of $a^3/\lambda^2$, see Fig.~\ref{fig:3}(a). Comparing with experimental data [blue symbols in Fig.~\ref{fig:3}(a)], it appears striking that early work with relatively short fibers \cite{Nagy} reported losses that deviate from predictions of our model whereas more recent reports with longer fibers \cite{Nagy2,Nagy3} appear to completely agree with our model. Moreover, we also compare the maximum beneficial length and the observed spectral broadening between theory and previous experimental findings in Fig.~\ref{fig:3}(b). Here it is not overly surprising that our projections are too optimistic, with maximum compressibilities and beneficial fiber lengths that are about two thirds of our predictions. Nevertheless, the relations derived from our spatial soliton model clearly explain the trends observed in previous experimentation, confirming that long hollow fibers promise superior performance compared to the single-meter-long segments of early experimentation.

In conclusion, the field of multimode nonlinear optics bears a number of appealing applications, which are mostly ruled by a $a^3/\lambda^2$ relationship. In particular, in hollow fibers, where group-velocity dispersion plays an inferior role, spatial soliton formation appears to take a previously unrecognized lead role at large diameters and lengths. Given that dispersive and absorptive lengths are orders of magnitude larger than the soliton length, adiabatic reshaping dominates the nonlinear dynamics of pulse propagation through the hollow fiber. As the critical power $P_{\rm cr}$ plays a decisive limiting role for the performance of a nonlinear multimode waveguide, further upscaling of compressible peak powers require usage of lower pressures or gases with lower refractive index than the commonly used argon. As the currently demonstrated highest peak powers already used fibers with several meter length, hosting even higher powers requires the use of significantly longer fibers of tens or even hundreds of meters length to accumulate sufficient nonlinear phase for the broadening process. Such dimensions appear to be out of range for universities but could certainly be implemented in large scale facilities, in particular linear accelerators. Using SCM fibers near their zero-dispersion wavelength instead, spectral broadening can be accomplished at much higher pulse energies than in single-mode fibers. In contrast to previous demonstrations of nonlinear multimode optics, our theoretical investigations suggest that the exact refractive index profile plays only a minor rule, enabling the use of simple step-index architectures rather than relying only on parabolic profiles.

\markboth{}{}

\vspace{8pt} \noindent {\bf Acknowledgments.} {\small GS gratefully acknowledges fruitful discussions with Pavel Sidorenko and Frank Wise (Cornell University) as well as with Howard Milchberg (UMD).}

\bibliography{MMsoliton_APS}

\begin{thebibliography}{42}%
\makeatletter
\providecommand \@ifxundefined [1]{%
 \@ifx{#1\undefined}
}%
\providecommand \@ifnum [1]{%
 \ifnum #1\expandafter \@firstoftwo
 \else \expandafter \@secondoftwo
 \fi
}%
\providecommand \@ifx [1]{%
 \ifx #1\expandafter \@firstoftwo
 \else \expandafter \@secondoftwo
 \fi
}%
\providecommand \natexlab [1]{#1}%
\providecommand \enquote  [1]{``#1''}%
\providecommand \bibnamefont  [1]{#1}%
\providecommand \bibfnamefont [1]{#1}%
\providecommand \citenamefont [1]{#1}%
\providecommand \href@noop [0]{\@secondoftwo}%
\providecommand \href [0]{\begingroup \@sanitize@url \@href}%
\providecommand \@href[1]{\@@startlink{#1}\@@href}%
\providecommand \@@href[1]{\endgroup#1\@@endlink}%
\providecommand \@sanitize@url [0]{\catcode `\\12\catcode `\$12\catcode
  `\&12\catcode `\#12\catcode `\^12\catcode `\_12\catcode `\%12\relax}%
\providecommand \@@startlink[1]{}%
\providecommand \@@endlink[0]{}%
\providecommand \url  [0]{\begingroup\@sanitize@url \@url }%
\providecommand \@url [1]{\endgroup\@href {#1}{\urlprefix }}%
\providecommand \urlprefix  [0]{URL }%
\providecommand \Eprint [0]{\href }%
\providecommand \doibase [0]{https://doi.org/}%
\providecommand \selectlanguage [0]{\@gobble}%
\providecommand \bibinfo  [0]{\@secondoftwo}%
\providecommand \bibfield  [0]{\@secondoftwo}%
\providecommand \translation [1]{[#1]}%
\providecommand \BibitemOpen [0]{}%
\providecommand \bibitemStop [0]{}%
\providecommand \bibitemNoStop [0]{.\EOS\space}%
\providecommand \EOS [0]{\spacefactor3000\relax}%
\providecommand \BibitemShut  [1]{\csname bibitem#1\endcsname}%
\let\auto@bib@innerbib\@empty
\bibitem [{\citenamefont {Chiao}\ \emph {et~al.}(1965)\citenamefont {Chiao},
  \citenamefont {Garmire},\ and\ \citenamefont {Townes}}]{Townes}%
  \BibitemOpen
  \bibfield  {author} {\bibinfo {author} {\bibfnamefont {R.~Y.}\ \bibnamefont
  {Chiao}}, \bibinfo {author} {\bibfnamefont {E.}~\bibnamefont {Garmire}},\
  and\ \bibinfo {author} {\bibfnamefont {C.~H.}\ \bibnamefont {Townes}},\
  }\bibfield  {title} {\bibinfo {title} {Self-trapping of optical beams},\
  }\href@noop {} {\bibfield  {journal} {\bibinfo  {journal} {Phys. Rev. Lett.}\
  }\textbf {\bibinfo {volume} {14}},\ \bibinfo {pages} {479} (\bibinfo {year}
  {1965})}\BibitemShut {NoStop}%
\bibitem [{\citenamefont {Zakharov}\ and\ \citenamefont
  {Shabat}(1972)}]{Zakharov}%
  \BibitemOpen
  \bibfield  {author} {\bibinfo {author} {\bibfnamefont {V.~E.}\ \bibnamefont
  {Zakharov}}\ and\ \bibinfo {author} {\bibfnamefont {A.~B.}\ \bibnamefont
  {Shabat}},\ }\bibfield  {title} {\bibinfo {title} {Exact theory of
  two-dimensional self-focusing and one-dimensional self-modulation of waves in
  nonlinear media},\ }\href@noop {} {\bibfield  {journal} {\bibinfo  {journal}
  {Sov. Phys. JETP}\ }\textbf {\bibinfo {volume} {34}},\ \bibinfo {pages} {62}
  (\bibinfo {year} {1972})}\BibitemShut {NoStop}%
\bibitem [{\citenamefont {Silberberg}(1990)}]{Silberberg}%
  \BibitemOpen
  \bibfield  {author} {\bibinfo {author} {\bibfnamefont {Y.}~\bibnamefont
  {Silberberg}},\ }\bibfield  {title} {\bibinfo {title} {Collapse of optical
  pulses},\ }\href@noop {} {\bibfield  {journal} {\bibinfo  {journal} {Opt.
  Lett.}\ }\textbf {\bibinfo {volume} {15}},\ \bibinfo {pages} {1282} (\bibinfo
  {year} {1990})}\BibitemShut {NoStop}%
\bibitem [{\citenamefont {Moll}\ \emph {et~al.}(2003)\citenamefont {Moll},
  \citenamefont {Gaeta},\ and\ \citenamefont {Fibich}}]{Moll}%
  \BibitemOpen
  \bibfield  {author} {\bibinfo {author} {\bibfnamefont {K.~D.}\ \bibnamefont
  {Moll}}, \bibinfo {author} {\bibfnamefont {A.~L.}\ \bibnamefont {Gaeta}},\
  and\ \bibinfo {author} {\bibfnamefont {G.}~\bibnamefont {Fibich}},\
  }\bibfield  {title} {\bibinfo {title} {Self-similar optical wave collapse:
  {observation} of the {Townes} profile},\ }\href@noop {} {\bibfield  {journal}
  {\bibinfo  {journal} {Phys. Rev. Lett.}\ }\textbf {\bibinfo {volume} {90}},\
  \bibinfo {pages} {203902} (\bibinfo {year} {2003})}\BibitemShut {NoStop}%
\bibitem [{\citenamefont {Majus}\ \emph {et~al.}(2014)\citenamefont {Majus},
  \citenamefont {Tamošauskas}, \citenamefont {Gražulevičiūtė},
  \citenamefont {Garejev}, \citenamefont {Lotti}, \citenamefont {Couairon},
  \citenamefont {Faccio},\ and\ \citenamefont {Dubietis}}]{Dubietis}%
  \BibitemOpen
  \bibfield  {author} {\bibinfo {author} {\bibfnamefont {D.}~\bibnamefont
  {Majus}}, \bibinfo {author} {\bibfnamefont {G.}~\bibnamefont {Tamošauskas}},
  \bibinfo {author} {\bibfnamefont {I.}~\bibnamefont {Gražulevičiūtė}},
  \bibinfo {author} {\bibfnamefont {N.}~\bibnamefont {Garejev}}, \bibinfo
  {author} {\bibfnamefont {A.}~\bibnamefont {Lotti}}, \bibinfo {author}
  {\bibfnamefont {A.}~\bibnamefont {Couairon}}, \bibinfo {author}
  {\bibfnamefont {D.}~\bibnamefont {Faccio}},\ and\ \bibinfo {author}
  {\bibfnamefont {A.}~\bibnamefont {Dubietis}},\ }\bibfield  {title} {\bibinfo
  {title} {Nature of spatiotemporal light bullets in bulk {Kerr} media},\
  }\href@noop {} {\bibfield  {journal} {\bibinfo  {journal} {Phys. Rev. Lett.}\
  }\textbf {\bibinfo {volume} {112}},\ \bibinfo {pages} {193901} (\bibinfo
  {year} {2014})}\BibitemShut {NoStop}%
\bibitem [{\citenamefont {Bespalov}\ and\ \citenamefont
  {Talanov}(1966)}]{Bespalov}%
  \BibitemOpen
  \bibfield  {author} {\bibinfo {author} {\bibfnamefont {V.~I.}\ \bibnamefont
  {Bespalov}}\ and\ \bibinfo {author} {\bibfnamefont {V.~I.}\ \bibnamefont
  {Talanov}},\ }\bibfield  {title} {\bibinfo {title} {Filamentary structure of
  light beams in nonlinear liquids},\ }\href@noop {} {\bibfield  {journal}
  {\bibinfo  {journal} {JETP Lett.}\ }\textbf {\bibinfo {volume} {3}},\
  \bibinfo {pages} {307} (\bibinfo {year} {1966})}\BibitemShut {NoStop}%
\bibitem [{\citenamefont {Bliss}\ \emph {et~al.}(1974)\citenamefont {Bliss},
  \citenamefont {Speck}, \citenamefont {Holzrichter}, \citenamefont {Erkkila},\
  and\ \citenamefont {Glass}}]{Bliss}%
  \BibitemOpen
  \bibfield  {author} {\bibinfo {author} {\bibfnamefont {E.~S.}\ \bibnamefont
  {Bliss}}, \bibinfo {author} {\bibfnamefont {D.~R.}\ \bibnamefont {Speck}},
  \bibinfo {author} {\bibfnamefont {J.~F.}\ \bibnamefont {Holzrichter}},
  \bibinfo {author} {\bibfnamefont {J.~H.}\ \bibnamefont {Erkkila}},\ and\
  \bibinfo {author} {\bibfnamefont {A.~J.}\ \bibnamefont {Glass}},\ }\bibfield
  {title} {\bibinfo {title} {Propagation of a high-intensity laser pulsewith
  small-scale intensity modulation},\ }\href@noop {} {\bibfield  {journal}
  {\bibinfo  {journal} {Appl. Phys. Lett.}\ }\textbf {\bibinfo {volume} {25}},\
  \bibinfo {pages} {448} (\bibinfo {year} {1974})}\BibitemShut {NoStop}%
\bibitem [{\citenamefont {Marburger}(1975)}]{Marburger}%
  \BibitemOpen
  \bibfield  {author} {\bibinfo {author} {\bibfnamefont {J.}~\bibnamefont
  {Marburger}},\ }\bibfield  {title} {\bibinfo {title} {Self-focusing:
  Theory},\ }\href@noop {} {\bibfield  {journal} {\bibinfo  {journal} {Progr.
  Quantum Electron.}\ }\textbf {\bibinfo {volume} {4}},\ \bibinfo {pages} {35}
  (\bibinfo {year} {1975})}\BibitemShut {NoStop}%
\bibitem [{\citenamefont {Boyd}(2020)}]{Boyd}%
  \BibitemOpen
  \bibfield  {author} {\bibinfo {author} {\bibfnamefont {R.~W.}\ \bibnamefont
  {Boyd}},\ }\href@noop {} {\emph {\bibinfo {title} {Nonlinear Optics}}},\
  \bibinfo {edition} {4th}\ ed.\ (\bibinfo  {publisher} {Academic Press,
  London, UK},\ \bibinfo {year} {2020})\BibitemShut {NoStop}%
\bibitem [{\citenamefont {Marcatili}\ and\ \citenamefont
  {Schmeltzer}(1964)}]{MarcatiliSchmeltzer}%
  \BibitemOpen
  \bibfield  {author} {\bibinfo {author} {\bibfnamefont {E.}~\bibnamefont
  {Marcatili}}\ and\ \bibinfo {author} {\bibfnamefont {R.}~\bibnamefont
  {Schmeltzer}},\ }\bibfield  {title} {\bibinfo {title} {Hollow metallic and
  dielectric waveguides for long distance optical transmission and lasers},\
  }\href@noop {} {\bibfield  {journal} {\bibinfo  {journal} {Bell System Tech.
  J.}\ }\textbf {\bibinfo {volume} {43}},\ \bibinfo {pages} {1783 } (\bibinfo
  {year} {1964})}\BibitemShut {NoStop}%
\bibitem [{\citenamefont {Nisoli}\ \emph {et~al.}(1996)\citenamefont {Nisoli},
  \citenamefont {{De Silvestri}},\ and\ \citenamefont {Svelto}}]{Nisoli}%
  \BibitemOpen
  \bibfield  {author} {\bibinfo {author} {\bibfnamefont {M.}~\bibnamefont
  {Nisoli}}, \bibinfo {author} {\bibfnamefont {S.}~\bibnamefont {{De
  Silvestri}}},\ and\ \bibinfo {author} {\bibfnamefont {O.}~\bibnamefont
  {Svelto}},\ }\bibfield  {title} {\bibinfo {title} {Generation of high energy
  10 fs pulses by a new pulse compression technique},\ }\href@noop {}
  {\bibfield  {journal} {\bibinfo  {journal} {Appl. Phys. Lett.}\ }\textbf
  {\bibinfo {volume} {68}},\ \bibinfo {pages} {2793} (\bibinfo {year}
  {1996})}\BibitemShut {NoStop}%
\bibitem [{\citenamefont {Nisoli}\ \emph {et~al.}(1997)\citenamefont {Nisoli},
  \citenamefont {{De Silvestri}}, \citenamefont {Svelto}, \citenamefont
  {Szipöcs}, \citenamefont {Ferenz}, \citenamefont {Spielmann}, \citenamefont
  {Sartania},\ and\ \citenamefont {Krausz}}]{Nisoli2}%
  \BibitemOpen
  \bibfield  {author} {\bibinfo {author} {\bibfnamefont {M.}~\bibnamefont
  {Nisoli}}, \bibinfo {author} {\bibfnamefont {S.}~\bibnamefont {{De
  Silvestri}}}, \bibinfo {author} {\bibfnamefont {O.}~\bibnamefont {Svelto}},
  \bibinfo {author} {\bibfnamefont {R.}~\bibnamefont {Szipöcs}}, \bibinfo
  {author} {\bibfnamefont {K.}~\bibnamefont {Ferenz}}, \bibinfo {author}
  {\bibfnamefont {C.}~\bibnamefont {Spielmann}}, \bibinfo {author}
  {\bibfnamefont {S.}~\bibnamefont {Sartania}},\ and\ \bibinfo {author}
  {\bibfnamefont {F.}~\bibnamefont {Krausz}},\ }\bibfield  {title} {\bibinfo
  {title} {Compression of high-energy laser pulses below 5 fs},\ }\href@noop {}
  {\bibfield  {journal} {\bibinfo  {journal} {Opt. Lett.}\ }\textbf {\bibinfo
  {volume} {22}},\ \bibinfo {pages} {522} (\bibinfo {year} {1997})}\BibitemShut
  {NoStop}%
\bibitem [{\citenamefont {Durfee}\ \emph {et~al.}(1999)\citenamefont {Durfee},
  \citenamefont {Rundquist}, \citenamefont {Backus}, \citenamefont {Herne},
  \citenamefont {Murnane},\ and\ \citenamefont {Kapteyn}}]{Durfee}%
  \BibitemOpen
  \bibfield  {author} {\bibinfo {author} {\bibfnamefont {C.~G.}\ \bibnamefont
  {Durfee}}, \bibinfo {author} {\bibfnamefont {A.~R.}\ \bibnamefont
  {Rundquist}}, \bibinfo {author} {\bibfnamefont {S.}~\bibnamefont {Backus}},
  \bibinfo {author} {\bibfnamefont {C.}~\bibnamefont {Herne}}, \bibinfo
  {author} {\bibfnamefont {M.~M.}\ \bibnamefont {Murnane}},\ and\ \bibinfo
  {author} {\bibfnamefont {H.~C.}\ \bibnamefont {Kapteyn}},\ }\bibfield
  {title} {\bibinfo {title} {Phase matching of high-order harmonics in hollow
  waveguides},\ }\href@noop {} {\bibfield  {journal} {\bibinfo  {journal}
  {Phys. Rev. Lett.}\ }\textbf {\bibinfo {volume} {83}},\ \bibinfo {pages}
  {2187} (\bibinfo {year} {1999})}\BibitemShut {NoStop}%
\bibitem [{\citenamefont {Skupin}\ \emph {et~al.}(2006)\citenamefont {Skupin},
  \citenamefont {Stibenz}, \citenamefont {Bergé}, \citenamefont {Lederer},
  \citenamefont {Sokollik}, \citenamefont {Schnürer}, \citenamefont
  {Zhavoronkov},\ and\ \citenamefont {Steinmeyer}}]{selfcompress}%
  \BibitemOpen
  \bibfield  {author} {\bibinfo {author} {\bibfnamefont {S.}~\bibnamefont
  {Skupin}}, \bibinfo {author} {\bibfnamefont {G.}~\bibnamefont {Stibenz}},
  \bibinfo {author} {\bibfnamefont {L.}~\bibnamefont {Bergé}}, \bibinfo
  {author} {\bibfnamefont {F.}~\bibnamefont {Lederer}}, \bibinfo {author}
  {\bibfnamefont {T.}~\bibnamefont {Sokollik}}, \bibinfo {author}
  {\bibfnamefont {M.}~\bibnamefont {Schnürer}}, \bibinfo {author}
  {\bibfnamefont {N.}~\bibnamefont {Zhavoronkov}},\ and\ \bibinfo {author}
  {\bibfnamefont {G.}~\bibnamefont {Steinmeyer}},\ }\bibfield  {title}
  {\bibinfo {title} {Self-compression by femtosecond pulse filamentation:
  Experiments versus numerical simulations},\ }\href@noop {} {\bibfield
  {journal} {\bibinfo  {journal} {Phys. Rev. E}\ }\textbf {\bibinfo {volume}
  {74}},\ \bibinfo {pages} {056604} (\bibinfo {year} {2006})}\BibitemShut
  {NoStop}%
\bibitem [{\citenamefont {Jargot}\ \emph {et~al.}(2018)\citenamefont {Jargot},
  \citenamefont {Daher}, \citenamefont {Lavenu}, \citenamefont {Delen},
  \citenamefont {N.~Forget},\ and\ \citenamefont {Georges}}]{multipass}%
  \BibitemOpen
  \bibfield  {author} {\bibinfo {author} {\bibfnamefont {G.}~\bibnamefont
  {Jargot}}, \bibinfo {author} {\bibfnamefont {N.}~\bibnamefont {Daher}},
  \bibinfo {author} {\bibfnamefont {L.}~\bibnamefont {Lavenu}}, \bibinfo
  {author} {\bibfnamefont {X.}~\bibnamefont {Delen}}, \bibinfo {author}
  {\bibfnamefont {M.~H.}\ \bibnamefont {N.~Forget}},\ and\ \bibinfo {author}
  {\bibfnamefont {P.}~\bibnamefont {Georges}},\ }\bibfield  {title} {\bibinfo
  {title} {Self-compression in a multipass cell},\ }\href@noop {} {\bibfield
  {journal} {\bibinfo  {journal} {Opt. Lett.}\ }\textbf {\bibinfo {volume}
  {43}},\ \bibinfo {pages} {5643} (\bibinfo {year} {2018})}\BibitemShut
  {NoStop}%
\bibitem [{\citenamefont {Sansone}\ \emph {et~al.}(2006)\citenamefont
  {Sansone}, \citenamefont {Benedetti}, \citenamefont {Calegari}, \citenamefont
  {Vozzi}, \citenamefont {Avaldi}, \citenamefont {Flammini}, \citenamefont
  {Poletto}, \citenamefont {Villoresi}, \citenamefont {Altucci}, \citenamefont
  {Velotta}, \citenamefont {Stagira}, \citenamefont {{De Silvestri}},\ and\
  \citenamefont {Nisoli}}]{Sansone}%
  \BibitemOpen
  \bibfield  {author} {\bibinfo {author} {\bibfnamefont {G.}~\bibnamefont
  {Sansone}}, \bibinfo {author} {\bibfnamefont {E.}~\bibnamefont {Benedetti}},
  \bibinfo {author} {\bibfnamefont {F.}~\bibnamefont {Calegari}}, \bibinfo
  {author} {\bibfnamefont {C.}~\bibnamefont {Vozzi}}, \bibinfo {author}
  {\bibfnamefont {L.}~\bibnamefont {Avaldi}}, \bibinfo {author} {\bibfnamefont
  {R.}~\bibnamefont {Flammini}}, \bibinfo {author} {\bibfnamefont
  {L.}~\bibnamefont {Poletto}}, \bibinfo {author} {\bibfnamefont
  {P.}~\bibnamefont {Villoresi}}, \bibinfo {author} {\bibfnamefont
  {C.}~\bibnamefont {Altucci}}, \bibinfo {author} {\bibfnamefont
  {R.}~\bibnamefont {Velotta}}, \bibinfo {author} {\bibfnamefont
  {S.}~\bibnamefont {Stagira}}, \bibinfo {author} {\bibfnamefont
  {S.}~\bibnamefont {{De Silvestri}}},\ and\ \bibinfo {author} {\bibfnamefont
  {M.}~\bibnamefont {Nisoli}},\ }\bibfield  {title} {\bibinfo {title} {Isolated
  single-cycle attosecond pulses},\ }\href@noop {} {\bibfield  {journal}
  {\bibinfo  {journal} {Science}\ }\textbf {\bibinfo {volume} {314}},\ \bibinfo
  {pages} {443} (\bibinfo {year} {2006})}\BibitemShut {NoStop}%
\bibitem [{\citenamefont {Ma}\ \emph {et~al.}(2016)\citenamefont {Ma},
  \citenamefont {Dostál},\ and\ \citenamefont {Brixner}}]{Ma}%
  \BibitemOpen
  \bibfield  {author} {\bibinfo {author} {\bibfnamefont {X.}~\bibnamefont
  {Ma}}, \bibinfo {author} {\bibfnamefont {J.}~\bibnamefont {Dostál}},\ and\
  \bibinfo {author} {\bibfnamefont {T.}~\bibnamefont {Brixner}},\ }\bibfield
  {title} {\bibinfo {title} {Broadband 7-fs diffractive-optic-based {2D}
  electronic spectroscopy using hollow-core fiber compression},\ }\href@noop {}
  {\bibfield  {journal} {\bibinfo  {journal} {Opt. Express}\ }\textbf {\bibinfo
  {volume} {18}},\ \bibinfo {pages} {20781} (\bibinfo {year}
  {2016})}\BibitemShut {NoStop}%
\bibitem [{\citenamefont {Nagy}\ \emph {et~al.}(2008)\citenamefont {Nagy},
  \citenamefont {Forster},\ and\ \citenamefont {Simon}}]{stretchedfiber}%
  \BibitemOpen
  \bibfield  {author} {\bibinfo {author} {\bibfnamefont {T.}~\bibnamefont
  {Nagy}}, \bibinfo {author} {\bibfnamefont {M.}~\bibnamefont {Forster}},\ and\
  \bibinfo {author} {\bibfnamefont {P.}~\bibnamefont {Simon}},\ }\bibfield
  {title} {\bibinfo {title} {Flexible hollow fiber for pulse compressors},\
  }\href@noop {} {\bibfield  {journal} {\bibinfo  {journal} {Appl. Opt.}\
  }\textbf {\bibinfo {volume} {47}},\ \bibinfo {pages} {3264} (\bibinfo {year}
  {2008})}\BibitemShut {NoStop}%
\bibitem [{\citenamefont {Nagy}\ \emph {et~al.}(2011)\citenamefont {Nagy},
  \citenamefont {Pervak},\ and\ \citenamefont {Simon}}]{Nagy}%
  \BibitemOpen
  \bibfield  {author} {\bibinfo {author} {\bibfnamefont {T.}~\bibnamefont
  {Nagy}}, \bibinfo {author} {\bibfnamefont {V.}~\bibnamefont {Pervak}},\ and\
  \bibinfo {author} {\bibfnamefont {P.}~\bibnamefont {Simon}},\ }\bibfield
  {title} {\bibinfo {title} {Optimal pulse compression in long hollow fibers},\
  }\href@noop {} {\bibfield  {journal} {\bibinfo  {journal} {Opt. Lett.}\
  }\textbf {\bibinfo {volume} {36}},\ \bibinfo {pages} {4422} (\bibinfo {year}
  {2011})}\BibitemShut {NoStop}%
\bibitem [{\citenamefont {Nagy}\ \emph {et~al.}(2019)\citenamefont {Nagy},
  \citenamefont {Hädrich}, \citenamefont {Simon}, \citenamefont {Blumenstein},
  \citenamefont {Walther}, \citenamefont {Klas}, \citenamefont {Buldt},
  \citenamefont {Stark}, \citenamefont {Breitkopf}, \citenamefont {Jójárt},
  \citenamefont {Seres}, \citenamefont {Várallyay}, \citenamefont {Eidam},\
  and\ \citenamefont {Limpert}}]{Nagy2}%
  \BibitemOpen
  \bibfield  {author} {\bibinfo {author} {\bibfnamefont {T.}~\bibnamefont
  {Nagy}}, \bibinfo {author} {\bibfnamefont {S.}~\bibnamefont {Hädrich}},
  \bibinfo {author} {\bibfnamefont {P.}~\bibnamefont {Simon}}, \bibinfo
  {author} {\bibfnamefont {A.}~\bibnamefont {Blumenstein}}, \bibinfo {author}
  {\bibfnamefont {N.}~\bibnamefont {Walther}}, \bibinfo {author} {\bibfnamefont
  {R.}~\bibnamefont {Klas}}, \bibinfo {author} {\bibfnamefont {J.}~\bibnamefont
  {Buldt}}, \bibinfo {author} {\bibfnamefont {H.}~\bibnamefont {Stark}},
  \bibinfo {author} {\bibfnamefont {S.}~\bibnamefont {Breitkopf}}, \bibinfo
  {author} {\bibfnamefont {P.}~\bibnamefont {Jójárt}}, \bibinfo {author}
  {\bibfnamefont {I.}~\bibnamefont {Seres}}, \bibinfo {author} {\bibfnamefont
  {Z.}~\bibnamefont {Várallyay}}, \bibinfo {author} {\bibfnamefont
  {T.}~\bibnamefont {Eidam}},\ and\ \bibinfo {author} {\bibfnamefont
  {J.}~\bibnamefont {Limpert}},\ }\bibfield  {title} {\bibinfo {title}
  {Generation of three-cycle multi-millijoule laser pulses at 318 {W} average
  power},\ }\href@noop {} {\bibfield  {journal} {\bibinfo  {journal} {Optica}\
  }\textbf {\bibinfo {volume} {6}},\ \bibinfo {pages} {1423} (\bibinfo {year}
  {2019})}\BibitemShut {NoStop}%
\bibitem [{\citenamefont {Nagy}\ \emph {et~al.}(2020)\citenamefont {Nagy},
  \citenamefont {Kretschmar}, \citenamefont {Vrakking},\ and\ \citenamefont
  {Rouzée}}]{Nagy3}%
  \BibitemOpen
  \bibfield  {author} {\bibinfo {author} {\bibfnamefont {T.}~\bibnamefont
  {Nagy}}, \bibinfo {author} {\bibfnamefont {M.}~\bibnamefont {Kretschmar}},
  \bibinfo {author} {\bibfnamefont {M.~J.~J.}\ \bibnamefont {Vrakking}},\ and\
  \bibinfo {author} {\bibfnamefont {A.}~\bibnamefont {Rouzée}},\ }\bibfield
  {title} {\bibinfo {title} {Generation of above-terawatt 1.5-cycle visible
  pulses at 1 {kHz} by post-compression in a hollow fiber},\ }\href@noop {}
  {\bibfield  {journal} {\bibinfo  {journal} {Opt. Lett.}\ }\textbf {\bibinfo
  {volume} {45}},\ \bibinfo {pages} {3313} (\bibinfo {year}
  {2020})}\BibitemShut {NoStop}%
\bibitem [{\citenamefont {Tempea}\ and\ \citenamefont {Brabec}(1998)}]{Tempea}%
  \BibitemOpen
  \bibfield  {author} {\bibinfo {author} {\bibfnamefont {G.}~\bibnamefont
  {Tempea}}\ and\ \bibinfo {author} {\bibfnamefont {T.}~\bibnamefont
  {Brabec}},\ }\bibfield  {title} {\bibinfo {title} {Theory of self-focusing in
  hollow waveguide},\ }\href@noop {} {\bibfield  {journal} {\bibinfo  {journal}
  {Opt. Lett.}\ }\textbf {\bibinfo {volume} {23}},\ \bibinfo {pages} {762}
  (\bibinfo {year} {1998})}\BibitemShut {NoStop}%
\bibitem [{\citenamefont {Fibich}\ and\ \citenamefont {Gaeta}(2000)}]{Fibich}%
  \BibitemOpen
  \bibfield  {author} {\bibinfo {author} {\bibfnamefont {G.}~\bibnamefont
  {Fibich}}\ and\ \bibinfo {author} {\bibfnamefont {A.}~\bibnamefont {Gaeta}},\
  }\bibfield  {title} {\bibinfo {title} {Critical power for self-focusing in
  bulk media and hollow waveguides},\ }\href@noop {} {\bibfield  {journal}
  {\bibinfo  {journal} {Opt. Lett.}\ }\textbf {\bibinfo {volume} {25}},\
  \bibinfo {pages} {335} (\bibinfo {year} {2000})}\BibitemShut {NoStop}%
\bibitem [{\citenamefont {Nurhuda}\ \emph {et~al.}(2003)\citenamefont
  {Nurhuda}, \citenamefont {Suda}, \citenamefont {Midorikawa}, \citenamefont
  {Hatayama},\ and\ \citenamefont {Nagasaka}}]{Nurhuda}%
  \BibitemOpen
  \bibfield  {author} {\bibinfo {author} {\bibfnamefont {M.}~\bibnamefont
  {Nurhuda}}, \bibinfo {author} {\bibfnamefont {A.}~\bibnamefont {Suda}},
  \bibinfo {author} {\bibfnamefont {K.}~\bibnamefont {Midorikawa}}, \bibinfo
  {author} {\bibfnamefont {M.}~\bibnamefont {Hatayama}},\ and\ \bibinfo
  {author} {\bibfnamefont {K.}~\bibnamefont {Nagasaka}},\ }\bibfield  {title}
  {\bibinfo {title} {Propagation dynamics of femtosecond laser pulses in a
  hollow fiber filled with argon: constant gas pressure versus differential gas
  pressure},\ }\href@noop {} {\bibfield  {journal} {\bibinfo  {journal} {J.
  Opt. Soc. Am. B}\ }\textbf {\bibinfo {volume} {20}},\ \bibinfo {pages} {2002}
  (\bibinfo {year} {2003})}\BibitemShut {NoStop}%
\bibitem [{\citenamefont {Andreasen}\ and\ \citenamefont
  {Kolesik}(2013)}]{Kolesik}%
  \BibitemOpen
  \bibfield  {author} {\bibinfo {author} {\bibfnamefont {J.}~\bibnamefont
  {Andreasen}}\ and\ \bibinfo {author} {\bibfnamefont {M.}~\bibnamefont
  {Kolesik}},\ }\bibfield  {title} {\bibinfo {title} {Midinfrared femtosecond
  laser pulse filamentation in hollow waveguides: A comparison of simulation
  methods},\ }\href@noop {} {\bibfield  {journal} {\bibinfo  {journal} {Phys.
  Rev. E}\ }\textbf {\bibinfo {volume} {87}},\ \bibinfo {pages} {053303}
  (\bibinfo {year} {2013})}\BibitemShut {NoStop}%
\bibitem [{\citenamefont {Safaei}\ \emph {et~al.}(2020)\citenamefont {Safaei},
  \citenamefont {Fan}, \citenamefont {Kwon}, \citenamefont {Légaré},
  \citenamefont {Lassonde}, \citenamefont {Schmidt}, \citenamefont {Ibrahim},\
  and\ \citenamefont {Légaré}}]{Bruno}%
  \BibitemOpen
  \bibfield  {author} {\bibinfo {author} {\bibfnamefont {R.}~\bibnamefont
  {Safaei}}, \bibinfo {author} {\bibfnamefont {G.}~\bibnamefont {Fan}},
  \bibinfo {author} {\bibfnamefont {O.}~\bibnamefont {Kwon}}, \bibinfo {author}
  {\bibfnamefont {K.}~\bibnamefont {Légaré}}, \bibinfo {author}
  {\bibfnamefont {P.}~\bibnamefont {Lassonde}}, \bibinfo {author}
  {\bibfnamefont {B.~E.}\ \bibnamefont {Schmidt}}, \bibinfo {author}
  {\bibfnamefont {H.}~\bibnamefont {Ibrahim}},\ and\ \bibinfo {author}
  {\bibfnamefont {F.}~\bibnamefont {Légaré}},\ }\bibfield  {title} {\bibinfo
  {title} {High-energy multidimensional solitary states in hollow-core
  fibres},\ }\href@noop {} {\bibfield  {journal} {\bibinfo  {journal} {Nature
  Photonics}\ }\textbf {\bibinfo {volume} {14}},\ \bibinfo {pages} {732}
  (\bibinfo {year} {2020})}\BibitemShut {NoStop}%
\bibitem [{\citenamefont {Crego}\ \emph {et~al.}(2019)\citenamefont {Crego},
  \citenamefont {Jarque},\ and\ \citenamefont {Roman}}]{Crego}%
  \BibitemOpen
  \bibfield  {author} {\bibinfo {author} {\bibfnamefont {A.}~\bibnamefont
  {Crego}}, \bibinfo {author} {\bibfnamefont {E.~C.}\ \bibnamefont {Jarque}},\
  and\ \bibinfo {author} {\bibfnamefont {J.~S.}\ \bibnamefont {Roman}},\
  }\bibfield  {title} {\bibinfo {title} {Influence of the spatial confinement
  on the self-focusing of ultrashort pulses in hollow-core fibers},\
  }\href@noop {} {\bibfield  {journal} {\bibinfo  {journal} {Sci. Rep.}\
  }\textbf {\bibinfo {volume} {9}},\ \bibinfo {pages} {9546} (\bibinfo {year}
  {2019})}\BibitemShut {NoStop}%
\bibitem [{\citenamefont {López-Zubieta}\ \emph {et~al.}(2018)\citenamefont
  {López-Zubieta}, \citenamefont {Jarque}, \citenamefont {Sola},\ and\
  \citenamefont {Roman}}]{Continuum}%
  \BibitemOpen
  \bibfield  {author} {\bibinfo {author} {\bibfnamefont {B.~A.}\ \bibnamefont
  {López-Zubieta}}, \bibinfo {author} {\bibfnamefont {E.~C.}\ \bibnamefont
  {Jarque}}, \bibinfo {author} {\bibfnamefont {{\' I}.~J.}\ \bibnamefont
  {Sola}},\ and\ \bibinfo {author} {\bibfnamefont {J.~S.}\ \bibnamefont
  {Roman}},\ }\bibfield  {title} {\bibinfo {title} {Spatiotemporal-dressed
  optical solitons in hollow-core capillaries},\ }\href@noop {} {\bibfield
  {journal} {\bibinfo  {journal} {{OSA} Continuum}\ }\textbf {\bibinfo {volume}
  {1}},\ \bibinfo {pages} {930} (\bibinfo {year} {2018})}\BibitemShut {NoStop}%
\bibitem [{\citenamefont {Haus}(2000)}]{Haus}%
  \BibitemOpen
  \bibfield  {author} {\bibinfo {author} {\bibfnamefont {H.~A.}\ \bibnamefont
  {Haus}},\ }\bibfield  {title} {\bibinfo {title} {Mode-locking of lasers},\
  }\href@noop {} {\bibfield  {journal} {\bibinfo  {journal} {IEEE J. Sel. Top.
  Quantum Electron.}\ }\textbf {\bibinfo {volume} {6}},\ \bibinfo {pages}
  {1173} (\bibinfo {year} {2000})}\BibitemShut {NoStop}%
\bibitem [{\citenamefont {Escoto}\ \emph {et~al.}(2021)\citenamefont {Escoto},
  \citenamefont {Demircan},\ and\ \citenamefont {Steinmeyer}}]{CageSoliton}%
  \BibitemOpen
  \bibfield  {author} {\bibinfo {author} {\bibfnamefont {E.}~\bibnamefont
  {Escoto}}, \bibinfo {author} {\bibfnamefont {A.}~\bibnamefont {Demircan}},\
  and\ \bibinfo {author} {\bibfnamefont {G.}~\bibnamefont {Steinmeyer}},\
  }\bibfield  {title} {\bibinfo {title} {Cage solitons},\ }\href@noop {}
  {\bibfield  {journal} {\bibinfo  {journal} {IEEE J. Quantum Electron.}\
  }\textbf {\bibinfo {volume} {57}},\ \bibinfo {pages} {1300106} (\bibinfo
  {year} {2021})}\BibitemShut {NoStop}%
\bibitem [{\citenamefont {Wise}(2010)}]{Wise}%
  \BibitemOpen
  \bibfield  {author} {\bibinfo {author} {\bibfnamefont {F.~W.}\ \bibnamefont
  {Wise}},\ }\bibfield  {title} {\bibinfo {title} {Generation of light
  bullets},\ }\href@noop {} {\bibfield  {journal} {\bibinfo  {journal}
  {Physics}\ }\textbf {\bibinfo {volume} {3}},\ \bibinfo {pages} {107}
  (\bibinfo {year} {2010})}\BibitemShut {NoStop}%
\bibitem [{\citenamefont {Renninger}\ and\ \citenamefont {Wise}(2013)}]{Wise2}%
  \BibitemOpen
  \bibfield  {author} {\bibinfo {author} {\bibfnamefont {W.~H.}\ \bibnamefont
  {Renninger}}\ and\ \bibinfo {author} {\bibfnamefont {F.~W.}\ \bibnamefont
  {Wise}},\ }\bibfield  {title} {\bibinfo {title} {Optical solitons in
  graded-index multimode fibres},\ }\href@noop {} {\bibfield  {journal}
  {\bibinfo  {journal} {Nature Communications}\ }\textbf {\bibinfo {volume}
  {4}},\ \bibinfo {pages} {1719} (\bibinfo {year} {2013})}\BibitemShut
  {NoStop}%
\bibitem [{\citenamefont {Lopez-Galmiche}\ \emph {et~al.}(2016)\citenamefont
  {Lopez-Galmiche}, \citenamefont {Eznaveh}, \citenamefont {Eftekhar},
  \citenamefont {Lopez}, \citenamefont {Wright}, \citenamefont {Wise},
  \citenamefont {Christodoulides},\ and\ \citenamefont
  {Correa}}]{Christodoulides}%
  \BibitemOpen
  \bibfield  {author} {\bibinfo {author} {\bibfnamefont {G.}~\bibnamefont
  {Lopez-Galmiche}}, \bibinfo {author} {\bibfnamefont {Z.~S.}\ \bibnamefont
  {Eznaveh}}, \bibinfo {author} {\bibfnamefont {M.~A.}\ \bibnamefont
  {Eftekhar}}, \bibinfo {author} {\bibfnamefont {J.~A.}\ \bibnamefont {Lopez}},
  \bibinfo {author} {\bibfnamefont {L.~G.}\ \bibnamefont {Wright}}, \bibinfo
  {author} {\bibfnamefont {F.}~\bibnamefont {Wise}}, \bibinfo {author}
  {\bibfnamefont {D.}~\bibnamefont {Christodoulides}},\ and\ \bibinfo {author}
  {\bibfnamefont {R.~A.}\ \bibnamefont {Correa}},\ }\bibfield  {title}
  {\bibinfo {title} {Visible supercontinuum generation in a graded index
  multimode fiber pumped at 1064 nm},\ }\href@noop {} {\bibfield  {journal}
  {\bibinfo  {journal} {Opt. Lett.}\ }\textbf {\bibinfo {volume} {41}},\
  \bibinfo {pages} {2553} (\bibinfo {year} {2016})}\BibitemShut {NoStop}%
\bibitem [{\citenamefont {Crenn}(1984)}]{Crenn}%
  \BibitemOpen
  \bibfield  {author} {\bibinfo {author} {\bibfnamefont {J.~P.}\ \bibnamefont
  {Crenn}},\ }\bibfield  {title} {\bibinfo {title} {Optical study of the
  {EH}$_{11}$ mode in a hollow circular oversized waveguide and {Gaussian}
  approximation of the far-field pattern},\ }\href@noop {} {\bibfield
  {journal} {\bibinfo  {journal} {Appl. Opt.}\ }\textbf {\bibinfo {volume}
  {23}},\ \bibinfo {pages} {3428} (\bibinfo {year} {1984})}\BibitemShut
  {NoStop}%
\bibitem [{\citenamefont {Goos}\ and\ \citenamefont
  {Hänchen}(1947)}]{GoosHaenchen}%
  \BibitemOpen
  \bibfield  {author} {\bibinfo {author} {\bibfnamefont {F.}~\bibnamefont
  {Goos}}\ and\ \bibinfo {author} {\bibfnamefont {H.}~\bibnamefont
  {Hänchen}},\ }\bibfield  {title} {\bibinfo {title} {Ein neuer und
  fundamentaler {Versuch} zur {Totalreflexion}},\ }\href@noop {} {\bibfield
  {journal} {\bibinfo  {journal} {Annalen der Physik}\ }\textbf {\bibinfo
  {volume} {436}},\ \bibinfo {pages} {333} (\bibinfo {year}
  {1947})}\BibitemShut {NoStop}%
\bibitem [{\citenamefont {Crenn}(1993)}]{Crenn2}%
  \BibitemOpen
  \bibfield  {author} {\bibinfo {author} {\bibfnamefont {J.~P.}\ \bibnamefont
  {Crenn}},\ }\bibfield  {title} {\bibinfo {title} {Optical propagation of the
  {HE}$_{11}$ mode and {Gaussian} beams in hollow circular waveguides},\
  }\href@noop {} {\bibfield  {journal} {\bibinfo  {journal} {Int. J. Infrared
  Millimeter Waves}\ }\textbf {\bibinfo {volume} {14}},\ \bibinfo {pages}
  {1947} (\bibinfo {year} {1993})}\BibitemShut {NoStop}%
\bibitem [{\citenamefont {Snyder}\ and\ \citenamefont
  {Love}(1983)}]{LoveSnyder}%
  \BibitemOpen
  \bibfield  {author} {\bibinfo {author} {\bibfnamefont {A.~W.}\ \bibnamefont
  {Snyder}}\ and\ \bibinfo {author} {\bibfnamefont {J.~D.}\ \bibnamefont
  {Love}},\ }\href@noop {} {\emph {\bibinfo {title} {Optical Waveguide
  Theory}}}\ (\bibinfo  {publisher} {Chapman and Hall, London, UK},\ \bibinfo
  {year} {1983})\BibitemShut {NoStop}%
\bibitem [{\citenamefont {Agrawal}(2019)}]{Agrawal}%
  \BibitemOpen
  \bibfield  {author} {\bibinfo {author} {\bibfnamefont {G.~P.}\ \bibnamefont
  {Agrawal}},\ }\href@noop {} {\emph {\bibinfo {title} {Nonlinear Fiber
  Optics}}},\ \bibinfo {edition} {6th}\ ed.\ (\bibinfo  {publisher} {Academic
  Press, London, UK},\ \bibinfo {year} {2019})\BibitemShut {NoStop}%
\bibitem [{\citenamefont {Vinetski\v{\i}}\ \emph {et~al.}(1979)\citenamefont
  {Vinetski\v{\i}}, \citenamefont {Kukhtarev}, \citenamefont {Odulov},\ and\
  \citenamefont {Soskin}}]{selfdiffract}%
  \BibitemOpen
  \bibfield  {author} {\bibinfo {author} {\bibfnamefont {V.~L.}\ \bibnamefont
  {Vinetski\v{\i}}}, \bibinfo {author} {\bibfnamefont {N.~V.}\ \bibnamefont
  {Kukhtarev}}, \bibinfo {author} {\bibfnamefont {S.~G.}\ \bibnamefont
  {Odulov}},\ and\ \bibinfo {author} {\bibfnamefont {M.~S.}\ \bibnamefont
  {Soskin}},\ }\bibfield  {title} {\bibinfo {title} {Dynamic self-diffraction
  of coherent light beams},\ }\href@noop {} {\bibfield  {journal} {\bibinfo
  {journal} {Sov. Phys. Usp.}\ }\textbf {\bibinfo {volume} {22}},\ \bibinfo
  {pages} {742} (\bibinfo {year} {1979})}\BibitemShut {NoStop}%
\bibitem [{\citenamefont {DeLong}\ \emph {et~al.}(1995)\citenamefont {DeLong},
  \citenamefont {Trebino},\ and\ \citenamefont {Kane}}]{Trebino}%
  \BibitemOpen
  \bibfield  {author} {\bibinfo {author} {\bibfnamefont {K.~W.}\ \bibnamefont
  {DeLong}}, \bibinfo {author} {\bibfnamefont {R.}~\bibnamefont {Trebino}},\
  and\ \bibinfo {author} {\bibfnamefont {D.}~\bibnamefont {Kane}},\ }\bibfield
  {title} {\bibinfo {title} {Comparison of ultrashort-pulse
  frequency-resolved-optical-gating traces for three common beam geometries},\
  }\href@noop {} {\bibfield  {journal} {\bibinfo  {journal} {J. Opt. Soc. Am.
  B}\ }\textbf {\bibinfo {volume} {11}},\ \bibinfo {pages} {1595} (\bibinfo
  {year} {1995})}\BibitemShut {NoStop}%
\bibitem [{\citenamefont {Chapman}\ \emph {et~al.}(2010)\citenamefont
  {Chapman}, \citenamefont {Butcher}, \citenamefont {Horak}, \citenamefont
  {Poletti}, \citenamefont {Frey},\ and\ \citenamefont {Brocklesby}}]{Chapman}%
  \BibitemOpen
  \bibfield  {author} {\bibinfo {author} {\bibfnamefont {R.~T.}\ \bibnamefont
  {Chapman}}, \bibinfo {author} {\bibfnamefont {T.~J.}\ \bibnamefont
  {Butcher}}, \bibinfo {author} {\bibfnamefont {P.}~\bibnamefont {Horak}},
  \bibinfo {author} {\bibfnamefont {F.}~\bibnamefont {Poletti}}, \bibinfo
  {author} {\bibfnamefont {J.~G.}\ \bibnamefont {Frey}},\ and\ \bibinfo
  {author} {\bibfnamefont {W.~S.}\ \bibnamefont {Brocklesby}},\ }\bibfield
  {title} {\bibinfo {title} {Modal effects on pump-pulse propagation in an
  {Ar}-filled capillary},\ }\href@noop {} {\bibfield  {journal} {\bibinfo
  {journal} {Opt. Express}\ }\textbf {\bibinfo {volume} {18}},\ \bibinfo
  {pages} {13279} (\bibinfo {year} {2010})}\BibitemShut {NoStop}%
\bibitem [{\citenamefont {Blow}\ \emph {et~al.}(1992)\citenamefont {Blow},
  \citenamefont {Doran},\ and\ \citenamefont {Phoenix}}]{solitonphase}%
  \BibitemOpen
  \bibfield  {author} {\bibinfo {author} {\bibfnamefont {K.~J.}\ \bibnamefont
  {Blow}}, \bibinfo {author} {\bibfnamefont {N.~J.}\ \bibnamefont {Doran}},\
  and\ \bibinfo {author} {\bibfnamefont {S.~J.~D.}\ \bibnamefont {Phoenix}},\
  }\bibfield  {title} {\bibinfo {title} {The soliton phase},\ }\href@noop {}
  {\bibfield  {journal} {\bibinfo  {journal} {Opt. Commun.}\ }\textbf {\bibinfo
  {volume} {88}},\ \bibinfo {pages} {137} (\bibinfo {year} {1992})}\BibitemShut
  {NoStop}%
\end{thebibliography}%

\end{document}